\documentclass[journal]{IEEEtran}

\usepackage{soul}
\usepackage{cite}
\usepackage{blindtext}
\usepackage{graphicx}
\usepackage{epsfig}
\usepackage{xcolor}
\usepackage{tabularx}
\usepackage{amssymb}
\usepackage{latexsym}
\usepackage{amsmath}
\usepackage{algorithm}
\usepackage[noend]{algpseudocode}
\usepackage{caption}
\usepackage{subcaption}
\usepackage[none]{hyphenat}
\usepackage{relsize}
\usepackage{amsthm}
\usepackage{tikz}
\newcommand*\circled[1]{\tikz[baseline=(char.base)]{
            \node[shape=circle,draw,inner sep=2pt] (char) {#1};}}

\begin{document}
%


\title{Prototyping Next Generation O-RAN Research Testbeds with SDRs\\
\vspace{10 pt}
\large{Integrating the Near-Real Time RIC, E2 Interface, and Open-Source Cellular Software}}

\author{Pratheek S. Upadhyaya$^1$,
        Aly Sabri Abdalla$^2$,
         Vuk Marojevic$^2$,
         Jeffrey H. Reed$^1$,
        and  Vijay K. Shah$^3$
        \\
        $^1$Bradley Department of Electrical and Computer Engineering, Virginia Tech, VA, USA \\
         $^2$Department of Electrical and Computer Engineering, Mississippi State University, MS, USA \\
        $^3$Cybersecurity Engineering Department, George Mason University, VA, USA \\
        Emails:\{pratheek,reedjh\}@vt.edu, \{asa298,vuk.marojevic\}@msstate.edu and vshah22@gmu.edu
}

\maketitle

\begin{abstract}
Open RAN (O-RAN) defines an emerging cellular radio access network (RAN) architecture for future 6G wireless networks, emphasizing openness and intelligence
While the inherent complexity and flexibility of the RAN gives rise to many new research problems, progress in developing solutions is hampered due to the lack of end-to-end, fully developed platforms that can help in pursuing use cases in realistic environments. This has motivated the formation of  open-source frameworks available to the wireless community. However, the rapid evolution of dedicated platforms and solutions utilizing various software-based technologies
often leave questions regarding the interoperability and interactions between the components in the framework. This article shows how to build a software-defined radio testbed featuring an open-source 5G 
system that can interact with the near-real-time (near-RT) RAN intelligent controller (RIC) of the O-RAN architecture through standard interfaces. 
We focus on the O-RAN E2 interface interactions, and outline the procedure to enable a RAN system with E2 capabilities. We demonstrate the working of two xApps on the testbed 
with detailed E2 message exchange procedures 
and their role in controlling next generation RANs. 
\end{abstract}

\section{Introduction} \label{intro}
The open radio access network, Open RAN or O-RAN, was 
introduced by the O-RAN Alliance---a consortium of industry and academic organizations---to enable true plug-and-play network solutions, with respect to both hardware and software~\cite{openranalliance}. Built on the two pillars of \textit{openness} and \textit{intelligence}, O-RAN focuses on breaking out from the closed vendor environment of legacy cellular networks and driving the mobile industry toward a multi-vendor and 
artificial intelligence (AI) empowered 
network architecture. This provides operators with more choices in programmability and customization of RAN elements, allowing them to mix and match 
network solutions with unprecedented network dynamics, reducing the capital and operational expenditures 
due to the influx of new entrants, and fostering faster innovations that improve end user quality of experience. In addition to openness, O-RAN also promotes disaggregation by decoupling network functions and harnessing software-defined networking (SDN) and network function virtualization (NFV) principles.

Applying data-driven learning approaches to network management 
is being regarded as the solution to deal with the increasing complexity of 5G and beyond wireless networks. The success achieved by machine learning (ML) for data processing and optimization has made it the perfect candidate to handle the demanding service requirements, constraints, stringent regulations, and rapidly changing scenarios. To this effect, the O-RAN architecture strives to embed intelligence in all layers of the network architecture and support cross-layer interactions by employing different RAN control loops~\cite{abdalla2021toward}. Moreover, access to closed RAN data to operators and 3rd parties along with vendors encourages deployment of services as RAN applications (xApps) that can leverage emerging ML techniques. 
This 
may lead to 
the first application store for next generation RANs (akin to the App store for Android). 

Since the past few years, interest in open-source platforms for cellular networks has been steadily growing. Attempts are being made by open-source software libraries such as srsRAN~\cite{srsRAN}, OpenAirInterface (OAI)~\cite{oai2021} to mitigate challenges like limited accessibility to actual cellular networks to members of the research community. These software bundles when used in conjunction with commercial software-defined radios (SDRs), offer quick instantiation of fully functional cellular network elements that comply to the standard and are compatible with commercial handsets~\cite{SDR}. With the introduction of O-RAN which runs mostly on \textit{whitebox servers}, the ease of development has increased manifold, allowing researchers to plan, prototype, analyze, and test their solutions in a real-world setting and leverage industry practices and expertise in the design of their solutions. 

Yet, there are issues to be dealt with when implementing the envisioned data-driven, intelligent, programmable, and virtualized networks. More clarity regarding the open source frameworks developed to bootstrap research efforts is required. Specifically, although the O-RAN Software Community (OSC)\footnote{https://wiki.o-ran-sc.org/display/ORAN} provides the code base for implementing the O-RAN architecture/interfaces 
and provides the RAN control stack, it does not offer the RAN nor does it directly interfaces with 
existing RAN implementations. The limited sample use cases 
and lack of systematic and holistic implementation solutions, processes, and support pose additional challenges. Other issues include missing details on exact functionalities and parameters controlled by each element, and the role of AI and data analytics in improving wireless network performance. 

 Recently, attempts have been made to alleviate these concerns and promote the sharing of knowledge among the O-RAN research community. Nikam et al.~\cite{niknam2020intelligent} 
 demonstrate AI controllers for O-RAN and present their integration into the O-RAN architecture. Balasubramanian et al.~\cite{balasubramanian2021ric} provide a detailed description of the O-RAN architecture implementation from a software perspective. Bonati et al.~\cite{bonati2020open} offer a survey of different O-RAN technologies including 
 RAN disaggregation, virtualization, and slicing. 
 Polese et al.~\cite{polese2021colo} introduce O-RAN system tools in conjunction with Colosseum, an SDR-based wireless channel emulator with LTE support. 
 Similarly, ~\cite{Johnson21NexRAN} reported a closed loop integration of O-RAN compliant software components for collecting data on the POWDER testbed with an open-source full-stack softwarized cellular network. 
 Both testbeds offer 4G experiments and provide open-source 
 solutions without providing the detailed design and integration of the core components for building an end-to-end O-RAN research platform. 
 
 This article describes one of the first works that attempts to bridge the gap between conceptualization and implementation 
 for enabling research on practical 6G technologies.
  We discuss the design requirements and implementation options while describing how to build an end-to-end wireless research testbed that is compliant with the O-RAN architecture and that enable research in a controlled, yet production-like setting. 
 We have open sourced the software\footnote{https://github.com/openaicellular/oaic}
 and provide comprehensive instructions and documentation\footnote{https://openaicellular.github.io/oaic/} to help the research community 
 recreating our testbed and
 gaining expertise utilizing these tools. 
 This is created with the hope that other researchers entering this field find it easier to understand the inner workings of the underlying software and hardware components. 

Section II gives a brief overview of O-RAN and srsRAN which serve as the base on which the testbed is developed. Section III discusses the testbed architecture and the employed software stack. In Section IV, we illustrate how srsRAN is enhanced with the O-RAN compliant E2 Agent and show how interactions can be established between the E2 Node and the near-real time RAN intelligent controller (near-RT RIC). Section V discusses use cases (xApps) and their implementation on our testbed. Finally, section VI concludes the paper. 

\section{Open-Source Cellular} \label{sec:background}

Recent years have seen significant open-source efforts  developing commercial-grade software-defined RAN solutions that can be used to set up standard-compliant cellular networks. 
Two popular software libraries are srsRAN and OpenAirInterface (OAI). Our testbed employs srsRAN 
since it is easily customizable, well documented, and supported. We outline the design choices  and explore the system configurations for each design choice.

\subsection{srsRAN}
The srsRAN software library is developed and maintained by Software Radio Systems (SRS). It implements 4G and 5G RAN and UE solutions fully in software with support for COTS SDR hardware. It consists of the 4G LTE/5G New Radio (NR) eNB/gNB, UE, and the 4G core network, or evolved packet core (EPC). The suite is also compatible with third-party mobile network solutions such as Amarisoft and Open5GS. srsRAN supports mobility management and inter-cell handovers \cite{KeithACM22}, among others. 


srsRAN is written in C/C++ and is distributed under AGPL v3 license. It is 
compatible with Ubuntu and Fedora Linux distributions. 
Moreover, the srsRAN provides a channel emulator in the downlink receive path to mimic uncorrelated fading channels, propagation delay and radio-Link failure. The 
channel emulator enables four different fading channel models which are single tap with no delay, extended pedestrian A (EPA), extended vehicular A (EVA), and extended typical urban (ETU) model as defined by the 3GPP. The software can be deployed on bare metal, virtual machines, or containers on a standard computer with 4 or more cores and at least 3 GHz clock. 
    
\paragraph{srsEPC}
The main functions provided by srsEPC are home subscriber service (HSS), mobility management entity (MME), service gateway (S-GW), and packet data network gateway (P-GW).  

\paragraph{srsENB}  
The srsENB functions include:    
\begin{itemize}
    \item 3GPP Release 10 for LTE and Release 15 support for 5G NSA,
    \item X2 interface for NSA, GTPU and standard S1AP interfaces to the core network, 
    \item Support for different transmission modes, 
    \item Frequency division duplex (FDD) with standard bandwidths of 1.4, 3, 5, 10, 15, and 20 MHz, 
    \item  Supports COTS SDRs, including USRPs, as well as ZeroMQ-based operation without radio hardware. 
\end{itemize}    

\paragraph{srsUE} The srsUE software 
implements the 4G/5G UE in software: 
\begin{itemize}
    \item Compliant with 3GPP Release 10 with features up to Release 15, supporting 5G NSA in FDD and TDD configurations,
    \item Enables evolved multimedia broadcast and multicast service (eMBMS), typically channel 
    equalizers, and a highly optimized Turbo decoder, 
    \item Supports different COTS hardware such as USRP B2x0 and X3x0, BladeRF, and LimeSDR, and ZeroMQ.
    \item Operates with soft USIM supporting XOR/Milenage authentication, hard USIM support via PC/SC, and snow3G and AES integrity/ciphering support.
\end{itemize} 

Two forms of execution are directly supported:
\begin{itemize}
    \item \textit{COTS or SDR UE Mode:} A LTE/5G COTS/SDR UE can be configured to access the 4G/5G network services by attaching to the en-gNB radio. When connected, the UE can either communicate locally or can access the Internet through an Internet-enabled EPC.
    \item \textit{Simulation Mode:} 
    srsRAN provides a simulated software environment to conduct experiments 
    with the help of the ZeroMQ networking library to transfer radio samples between applications without SDR/RF hardware. The communication between a UE and the network is established by creating separate network namespaces. 
\end{itemize}

\subsection{O-RAN}
\begin{figure}[t]
    \centering
    \includegraphics[width=0.49\textwidth]{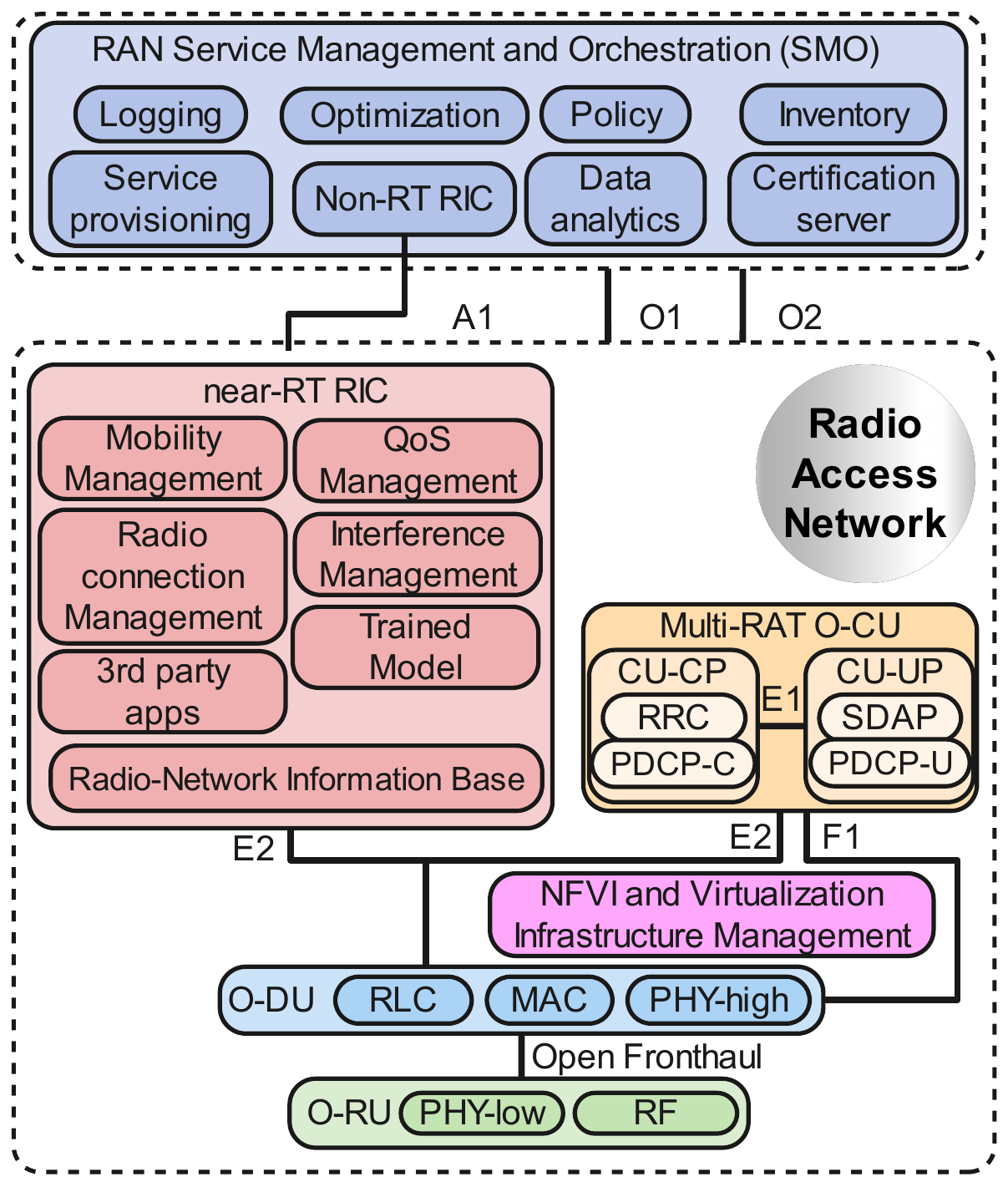}
    \caption{The O-RAN architecture. 
    } 
    \label{fig:ORANARCH}
\end{figure}
The realization of O-RAN’s vision is accomplished by adopting interfaces and operating software to separate the RAN control plane from the user plane and building a modular base station software stack that operates on 
COTS hardware. In the O-RAN architecture, the radio side includes the near-RT 
RIC, the O-RAN centralized unit (O-CU), the O-RAN distributed unit (O-DU), and the O-RAN radio unit (O-RU). The management side features the service management and orchestration (SMO) framework that contains non-RT RIC functions. Both the RAN and management components of the O-RAN architecture are illustrated in Fig.~\ref{fig:ORANARCH}.

By harnessing SDN and NFV principles, RAN disaggregation is introduced and the higher layers of the 3GPP protocol stack---radio resource control (RRC), service data adaption protocol (SDAP), and packet data convergence protocol (PDCP)---are split from the lower layers---radio link control (RLC), medium access control (MAC), and physical (PHY) layers. The two logical units hosting these functions are the O-CU and the O-DU. The lower part of the PHY (PHY-Low) in DU is further separated into a standalone O-RU, introducing the 7.2x functional split. 
The O-CU, located at the edge cloud and capable of controlling multiple O-DUs on a per UE basis, takes care of network functions such as call admission, UE connection, and traffic flow management while the DU (on-site eNB/gNB) is restricted to lower RAN layer functions such as scheduling, retransmission handling, and resource allocation. 
Disaggregation is taken to the next level in the O-CU by further dividing it into the logical control plane (CP), which hosts the RRC and control plane segment of the PDCP protocol, and one or more user planes (UPs), which manages the multiplexing of IP flows into logical gNB buffers and handles the SDAP and user plane part of the PDCP protocol. Moreover, the traditional CP is again partitioned into the O-CU-CP 
and the RIC. The RIC is a SDN-based component that embodies the network intelligence and performs selected radio resource management (RRM) functions.

Based on the operating timescales, the RIC is split into the near-RT RIC and the non-Real Time RIC. The near-RT RIC undertakes fine-grained data collection from the lower layers and uses it 
by software plugins called xApps to make RRM decisions. The non-RT RIC 
collects data from all the layers, processes the data by rApps, and provides policies and intent information to the xApps. 
Training, configuration, selection of ML models using the collected data and dispatching them to the near-RT RIC takes place in this segment.

To ensure proper communication between all these different protocol and management layers, open interfaces are standardized. Since all components are exposed to the same interface APIs, the substitution of components with better performing implementations is much easier than for legacy cellular networks. 
For the interactions between the O-CU and the O-DUs, the F1 (fronthaul) interface is based on the enhnced common public radio interface (eCPRI). The E1 interface is used for communication between the O-CU-CP and O-CU-UP. 3GPP has also defined the NG-c and NG-u interfaces to enable communication from O-CU-CP and O-CU-UP to 5G Core netork functions. 
The O-RAN Alliance introduces two additional control interfaces, the E2 and the operations \& administration (O1) interfaces. 
Through the E2 south-bound near-RT RIC interface, per user data from the RAN is provided 
to the xApps. 
The control decisions/policies determined by these xApps are propagated to the O-CU (CP/UP) or O-DU accordingly. Through the A1 north-bound near-RT RIC interfaces, 
policy guidance and information is provided to the xApps. 

O-RAN envisions three control loops operating at timescales that range from 1 ms to 1 s or more. 
Specifically, the non-RT RIC performs operations with a time granularity higher than 1 s, for model training
and service provisioning, among others. The near-RT RIC handles procedures involving timescales of 10 ms and above, 
implementing RAN intelligence through the near-RT data-driven control loop.

\section{O-RAN Compatible 5G NSA Testbed} 
\vspace{-1 mm}
\label{sec:testbed}
\begin{figure}[t]
    \centering
    \includegraphics[width=0.48\textwidth]{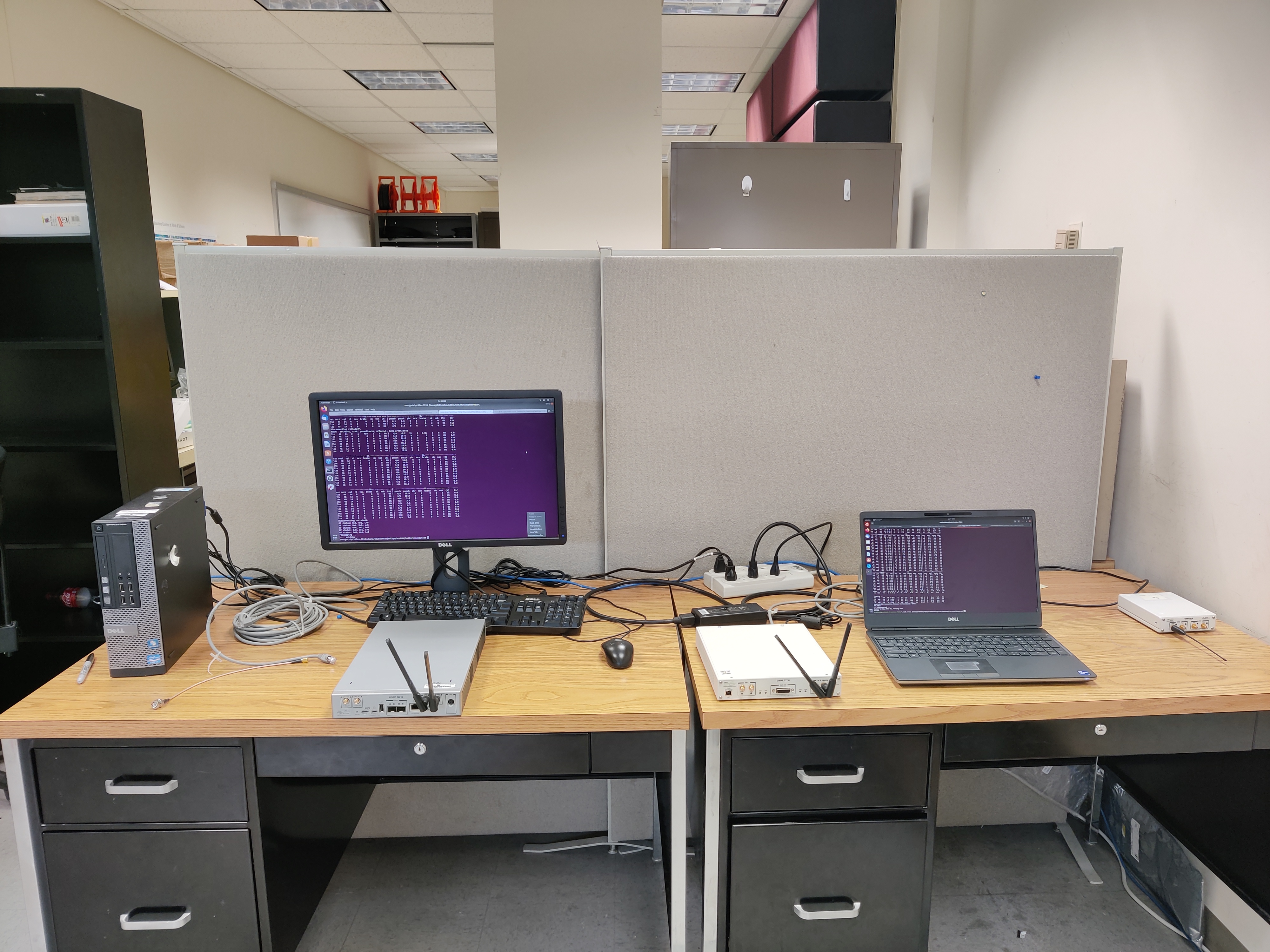}
    \caption{srsRAN 
    base station and core network (left) and UE (right) in a laboratory setup. 
    }
    \label{fig:testbed_srsran}
\end{figure}

Our testbed comprises of the srsRAN cellular stack and the near-RT RIC to establish the near-RT closed control loop. In this section, we consider the srsRAN and near-RT RIC deployment separately and discuss our primary design choices. 
\subsection{srsRAN Cellular Stack Deployment}

\subsubsection{Hardware Components}
srsRAN runs on commercial off-the-shelf (COTS) compute and RF hardware. 
Our testbed includes a total of 14 radio nodes. 
The Ettus Research Octoclock enables the synchronization or radios. 
Each compute node is a workstation featuring a 11th generation Intel Core-i9 processors with 24 MB of cache memory, 8 cores, and  up to 5 GHz per core. 
Each compute node supports the AVX-512 instruction set for increased performance. The radio equipment can be a LimeSDRs, SoapySDRs, or another SDR, but our testbed uses Ettus Research B205-mini, B-210, X-310, and N310, models. 
The specific hardware requirements are a function of the 4G/5G configuration. 
Fig. \ref{fig:testbed_srsran} shows our 5G testbed deployment using X310 USRP. A single en-gNB and a single UE in close proximity is depicted in the figure. Other USRPs are deployed across the laboratory. 

\subsubsection{Design choices}
The design choices are summarized in table \ref{tab:my_table} and discussed in continuation.
\begin{table}[htb]
\caption{Minimum hardware configuration for 4G \& 5G NSA network implementation.}
\label{tab:my_table} 
\includegraphics[width=0.49\textwidth]{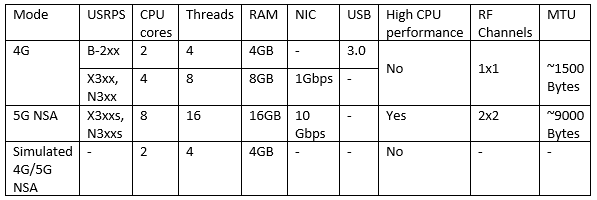}
\end{table}
\paragraph{LTE Mode} The general base station can use any of the SDR hardware mentioned previously. The B205-mini and B210 USRP connect via USB 3.0 to the host computer and require modest processing power and few cores. The data transfer between the compute node and the higher model USRPs (e.g., X310 or N310) can also be achieved through a standard 1 GigE. Both the eNB and UE use a single channel. The low system specification requirements enable the usage of 
virtual machines and Linux containers and can effectively execute on a 4-core Intel NUC10 computer. 

\paragraph{5G NSA Mode} The general Node B (gNB) 
requires an X310 or an N310 USRP connected to the compute node over a 10 Gbe link. 
The 10G link is essential to efficiently tranfer the high volume of I/Q samples from the compute node to the USRP and vice versa. Also, 5G NSA needs at least two channels, one for the 5G and one for the 4G RAN signaling. 
The compute node requirements 
are correspondingly higher than for the LTE-only mode. We recommend at least 8 physcial cores. 


\subsection{
RIC Deployment}
\subsubsection{Hardware}
We deploy the near-RT RIC interfaces and processes provided by OSC to interact with the SMO and the RAN. 
The near-RT RIC is deployed on a virtualized workstation with an Intel i9 11th Gen Processor, 64 GB of RAM, an Nvidia RTX A-5000 GPU to handle AI functionality, and 1 TB of storage capacity.
Virtualization is implemented at the operating system level using Virtual Infrastructure Managers such as Kubernetes. Each layer of the O-RAN stack implements virtualization through a microservice based architecture, where an application is decomposed into several parts running in their own lightweight environment. This makes the system more robust and makes it easier to address performance issues by creating multiple instances of the microservice. Scalability, agile testing and integration of upgrades, and independent applications are additional benefits.  

\subsubsection{O-RAN Software Community}
OSC in collaboration with the Linux Foundation contribute to open-source software that is compliant with the O-RAN specifications. Source code for multiple RAN control components is periodically released. 
Our testbed adopts the near-RT RIC software package with individual components being deployed as Kubernetes pods. Here, we briefly describe the different services handled by these components

\paragraph{xApps}
xApps are software applications that can leverage AI/ML based algorithms to optimize RAN functions. The xApps are the heart of the near-RT RIC. They reconfigure RAN functions and automatically collect RAN data reports. xApps can have contracts with RAN functions through service models (SMs) or they can act independently  for operations such as collecting metrics by interacting with the various components of the near-RT RIC. 

\paragraph{Interface Termination Points}
The interaction with RAN (E2 Nodes) goes through the E2 Termination.  The xApps interact with the SMO, non-RT RIC, and external servers for policy directives and enrichment information through the A1 Mediator. The RAN data, xApp decisions and metrics and other feedback is reported to the SMO and the non-RT RIC through the O1 Mediator. 


\paragraph{Supporting Components}
\begin{itemize}
    \item E2 Manager: helps in authenticating and registering E2 nodes in the near-RT RIC's database and maintains the status of the RAN's connection to the near-real time RIC.
    \item Shared Data Layer: stores RAN specific information in this persistent storage for consumption by xApps.
    \item Subscription service: provides subscription services 
    to E2 Nodes. 
    \item Routing Manager: handles all internal communications among the O-RAN components and 
    enables advanced functionalities such as xApp chaining. 
\end{itemize}




\section{E2 Workflow} 
\vspace{-1 mm}
\label{sec:workflow}
In this section, we demonstrate how the use of a common standardized interface can help in mixing and matching software and hardware components provided by multiple vendors. Specifically, we discuss the integration of the E2 agent with srsRAN and detail how standardized procedures can be incorporated to establish communication with the near-RT RIC. 

\subsection{E2 Interface Integration}
E2 is a logical point-to-point interface between the near-real time RIC and one or more eNB/gNBs, O-DUs, O-CU-CPs or O-CU-UPs, which are called \textit{E2 Nodes}. 
Multiple RAN configurations are supported depending on the E2 agent associated with the network elements. 
xApps 
can control and optimize cell level and UE level RAN functions through this interface. In the following sections we provide a brief overview of the E2 interface, the underlying protocol, and practical O-RAN implementation/integration principles. 

\subsubsection{
How Openness is achieved}
\begin{figure}[t]
    \centering
    \includegraphics[width=0.49\textwidth]{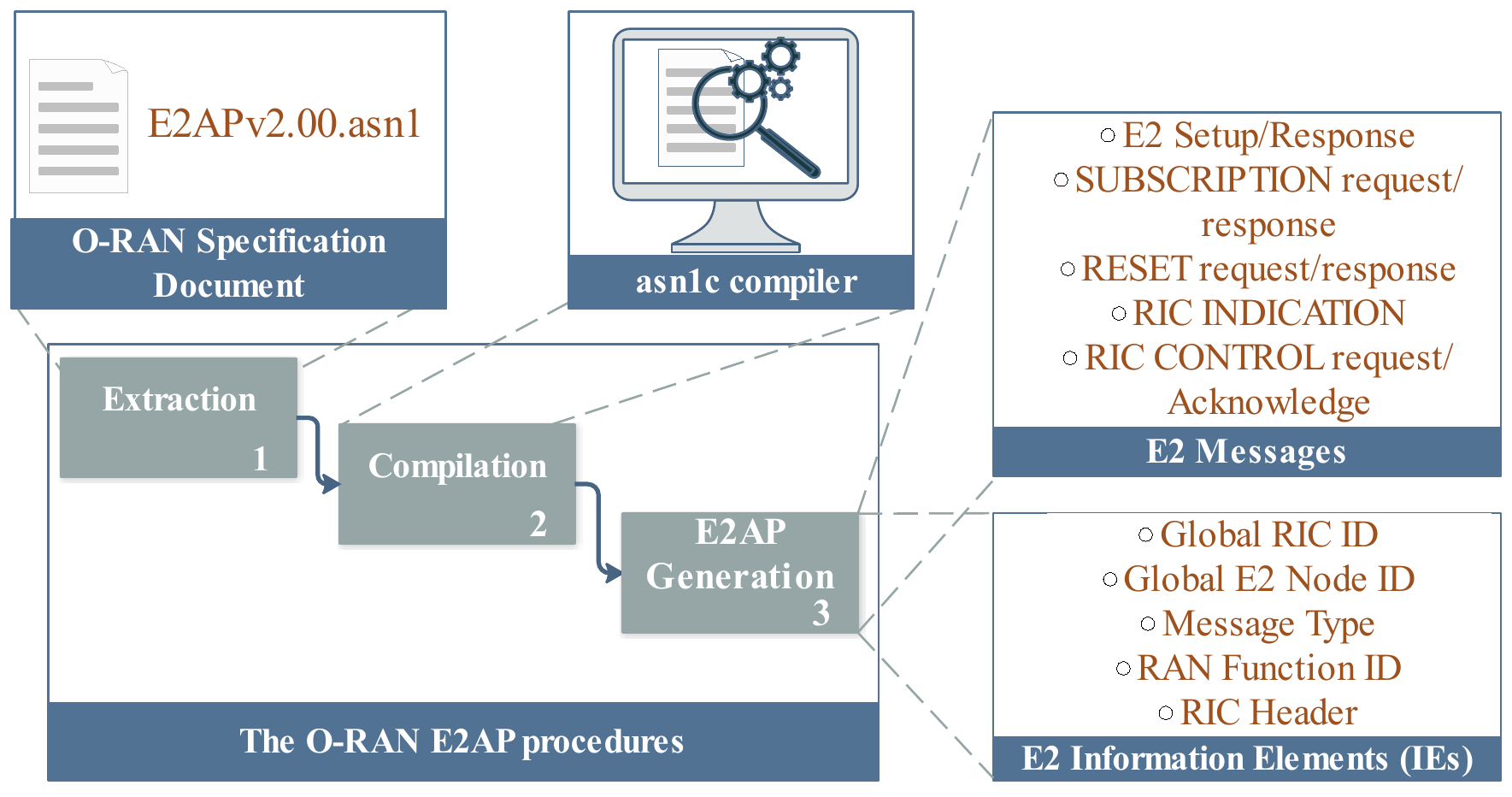}
    \caption{The O-RAN E2AP procedures.} 
    \label{fig:E2AP_asn1}
\end{figure}
The O-RAN Alliance Working Group 3 provides the specifications for the E2 related procedures and Information Elements (IEs) in the E2 Application Protocol (E2AP) document \cite{openranalliance}. The nomenclature 
is inspired by the Xn interface specifications of 3GPP with some elements being reused. E2 messages are comprised of one or more IEs which define the contents of the message. A sequential exchange of messages is called a procedure. E2 supports a variety of procedures divided into support and service class procedures. The support class procedures include interface management functions such as E2 Setup, E2 Reset and E2 Update, whereas the service class procedures support RAN CONTROL or REPORT functions.

The E2 procedure definitions, message format, and IE definitions are also provided in the Abstract Syntax Notation 1 (ASN.1) Style ~\cite{kaliski1993layman}. ASN.1 is a language agnostic interface description language for defining data structures used in standard protocols. This encoding establishes common interface definitions for two entities to communicate with each other. Each message or IE is a data structure that can be directly compiled into libraries that encode or decode the information to BER/XML formats. 
Even though implementations may vary, since the final encoding of the information is in ASN.1 compatible format, it can be decoded by any other system employing the same ASN.1 source specification. For example, the E2 SETUP request can be implemented in different ways by different vendors. But the near-RT RIC will face no issues in decoding the information since it employs the E2AP ASN.1 definitions. Fig. \ref{fig:E2AP_asn1} gives a schematic overview of the extraction and usage of these standard protocol definitions. In our integration of the E2 interface with srsRAN, we use the asn1c compiler to obtain the source code that allows us to encode and decode the E2 messages. 


\subsubsection{E2 Setup Procedure}
\begin{figure*}[t]
    \centering
    \includegraphics[width=0.99\textwidth]{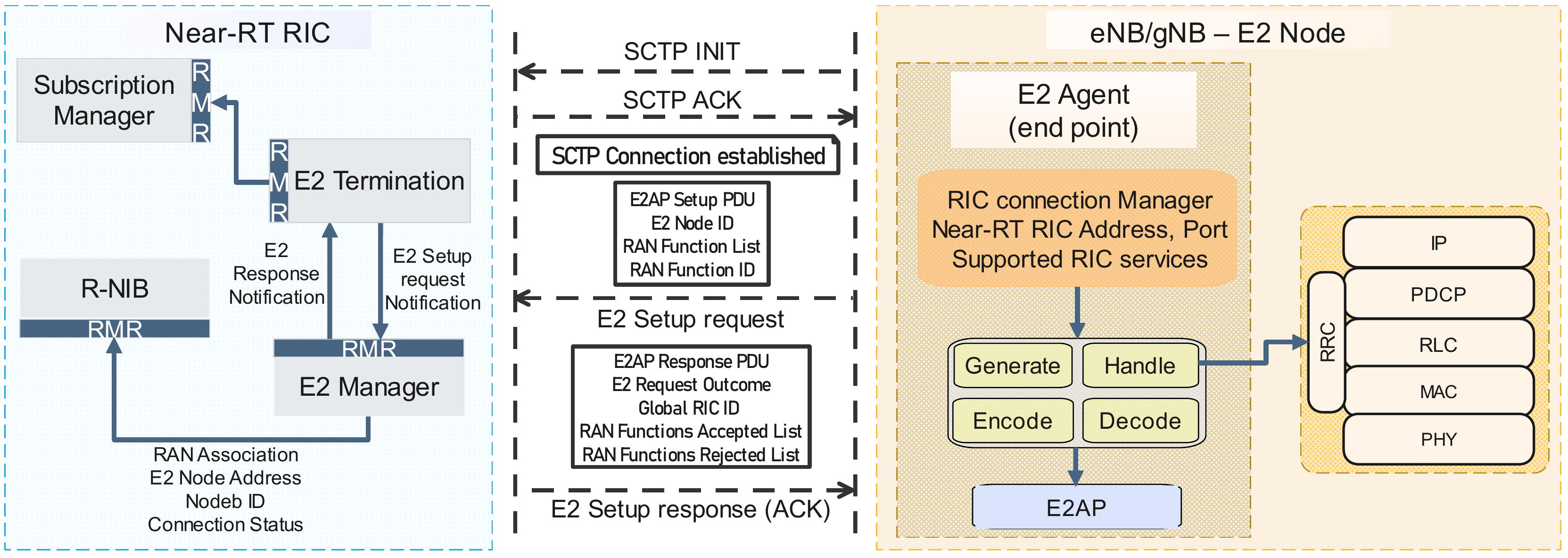}
    \caption{E2 Setup: illustration of the successful signalling procedure.}
    \label{fig:E2AP_setup}
\end{figure*}
The E2 agent is a separate entity from the RAN protocol stack and only interfaces with specific RAN functions. The agent associates with the near-RT RIC employing the stream control transmission protocol (SCTP). 
SCTP connections have been commonly used in LTE since it provides better security and features like multi-homing and multi-streaming\cite{shahdad2014multihoming}. Multi-streaming enables the exchange of E2 messages for different RAN functions independently across multiple streams, thus avoiding unnecessary head of line blocking as seen in TCP \cite{shahdad2014multihoming}. 

Fig. \ref{fig:E2AP_setup} details the successful E2 Setup procedure between a single E2 Node and the near-RT RIC. As soon as the eNB/gNB is up, it generates a E2 SETUP REQUEST message by including the RAN ID and the functions it supports as IEs. Here we use the source specifications for encoding purposes. The E2 termination at the near-RT RIC receives this request and forwards it to the E2 Manager, which decodes the request and performs authentication procedures. The RAN IP, ID, and connection status are maintained in the Radio Network Information Database (R-NIB). 


\section{xApp Deployment and Performance Analysis} 
\vspace{-1 mm}
\label{sec:usecases}
In this section, we describe the configuration of E2 Nodes to establish service related communication with the RIC. We then show through example use cases (xApps), the design approaches employed to achieve this in our testbed. 


\subsection{RAN Services \& E2 Service Models}
The E2AP protocol enables configuration of each E2 Node to support RIC services (reporting measurements and metrics for use within the RIC and/or for RIC control of RAN parameters). Each RAN function in an E2 Node can make use of an individual or a combination of RIC services. Currently the E2AP supports four main RIC services: REPORT, INSERT, CONTROL and POLICY ~\cite{polese2022understanding}. 
Each RAN function using these services has the procedures, message types, formats, and IE defined in its own E2 SM using ASN.1 encoding. The RAN functions that are exposed to the RIC are relayed during the E2 Setup.

xApps need to have the same SM definitions to effectively communicate (encode and decode E2 messages) and control RAN parameters. Therefore, SMs can be viewed as a form of contract between xApps and RAN functions. 

\subsection{xApp Deployment and Subscription Procedure}
During the onboarding of an xApp 
a descriptor file identifies the type of messages it consumes and generates. Once deployed, the xApp initiates a Subscription Procedure to find the E2 Nodes supporting the RIC services it offers. The RIC subscription is an E2AP procedure that also installs event triggers and subsequent service actions. 
Event triggers specify when an E2 node should send E2 messages to the RIC. An event trigger can be some event within the RAN or 
a periodic timer. When the timer expires, an indication message is sent by the E2 node.

\subsection{Lifecycle of the KPIMON xApp}
The first xApp that we deploy is a modified version of the KPIMON xApp provided by OSC. This xApp is based on the E2SM-KPM v1.01 SM \cite{openranalliance} and collects key performance indicators (KPIs) from the RAN. 


\begin{figure*}[t]
    \centering
    \includegraphics[width=0.9\textwidth]{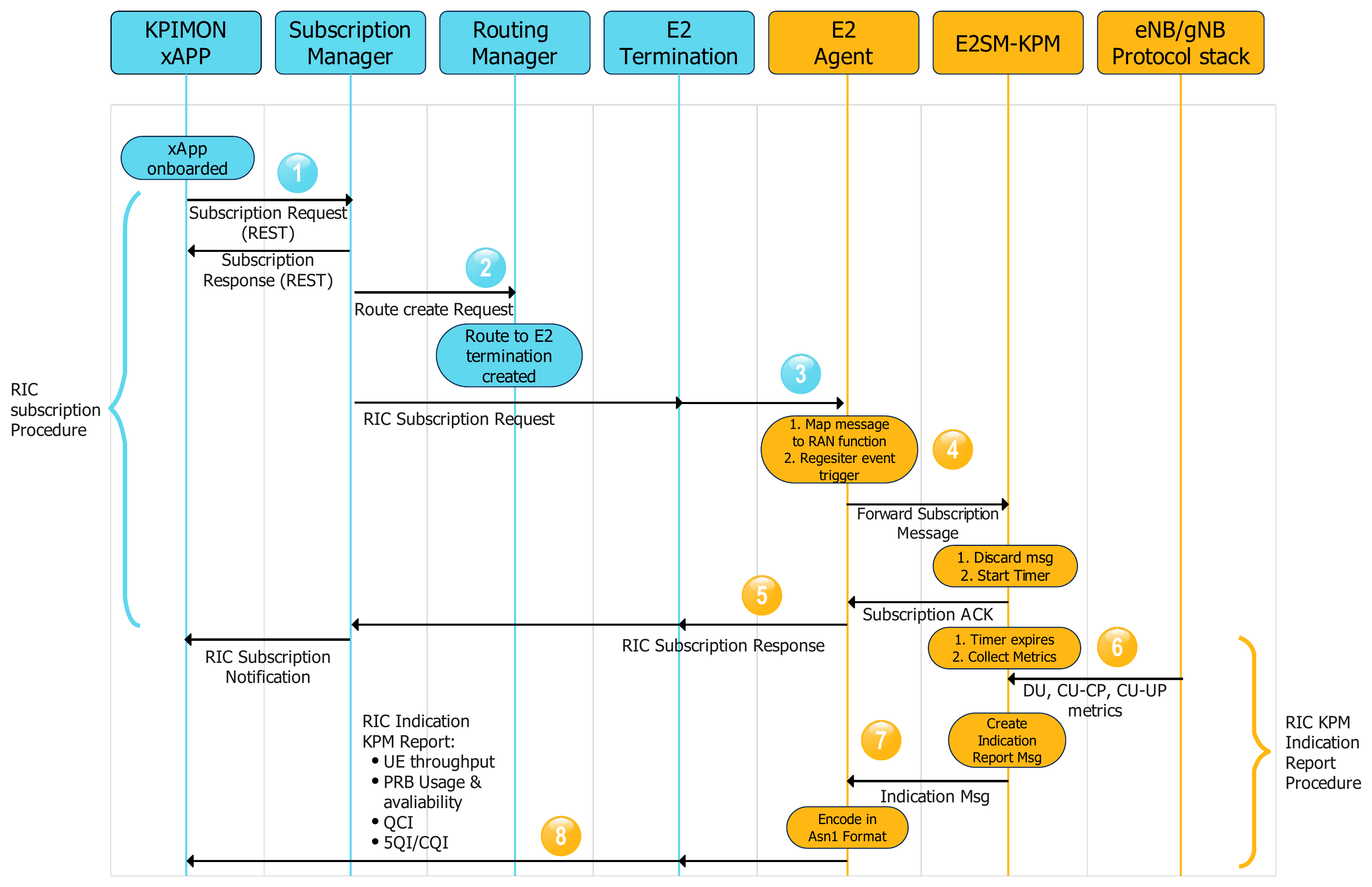}
    \caption{End-to-end workflow and signaling procedure for the collection of KPIs. The RIC Subscription and RIC Indication Procedures are used by the KPIMON xApp.}
    \label{fig:KPIMON_end_to_end}
\end{figure*}
Fig. \ref{fig:KPIMON_end_to_end} details the end-to-end signalling procedure, starting with the initiation of the subscription procedure and terminating with the reception of indication messages containing the KPM reports. \circled{1} When the xApp is deployed, it communicates with the Subscription Manager to initiate a subscription request procedure to all E2 Nodes (based on their IDs) that support the functionality that the xApp offers. Note that the E2 Node would have already indicated the RAN functions' supporting RIC services and the corresponding SMs during the E2 Setup procedure. \circled{2} This RIC subscription request is routed to the E2 Termination. 
\circled{3} The E2 Termination forwards it to 
the E2 Agent at the E2 Node. \circled{4} The E2 Agent decodes the RAN Function ID and service from the message, maps it to the KPM RAN function and forwards the subscription message to E2SM-KPM agent for further processing. 
\circled{5} The E2SM-KPM agent decodes the subscription message to determine the periodicity of reports and initiates a timer. Then, a RIC subscription response notifies the xApp about successful subscription. \circled{6} Upon expiration of the timer at the E2 Node, the O-DU, O-CU-CP \& O-CU-UP metrics for the previous time period are collected by the E2-SM KPM agent. The metrics collected from each of these network elements are packaged in containers, where each container has its unique ID. Each E2 message contains a header with a global RAN ID (PLMN ID) to identify the RAN node. 
\circled{7} These metrics are compiled to form the indication message (KPM Report) by using the appropriate IEs defined in the SM and encoded using ASN1 encoding. \circled{8} The indication messages are sent periodically to the KPIMON xApp. 
The xApp uses the same SM and Asn1 definitions to decode the indication message and obtain the metrics. The metrics collected by KPIMON include the number of utilized and available Physical Resource Blocks (PRBs), the number of active UE connections, QoS Class Identifiers (QCI) with active data transmissions, and the downlink and uplink data transferred in bytes in each session for each QCI. 

\begin{figure}[t]
    \centering
    \includegraphics[width=0.49\textwidth]{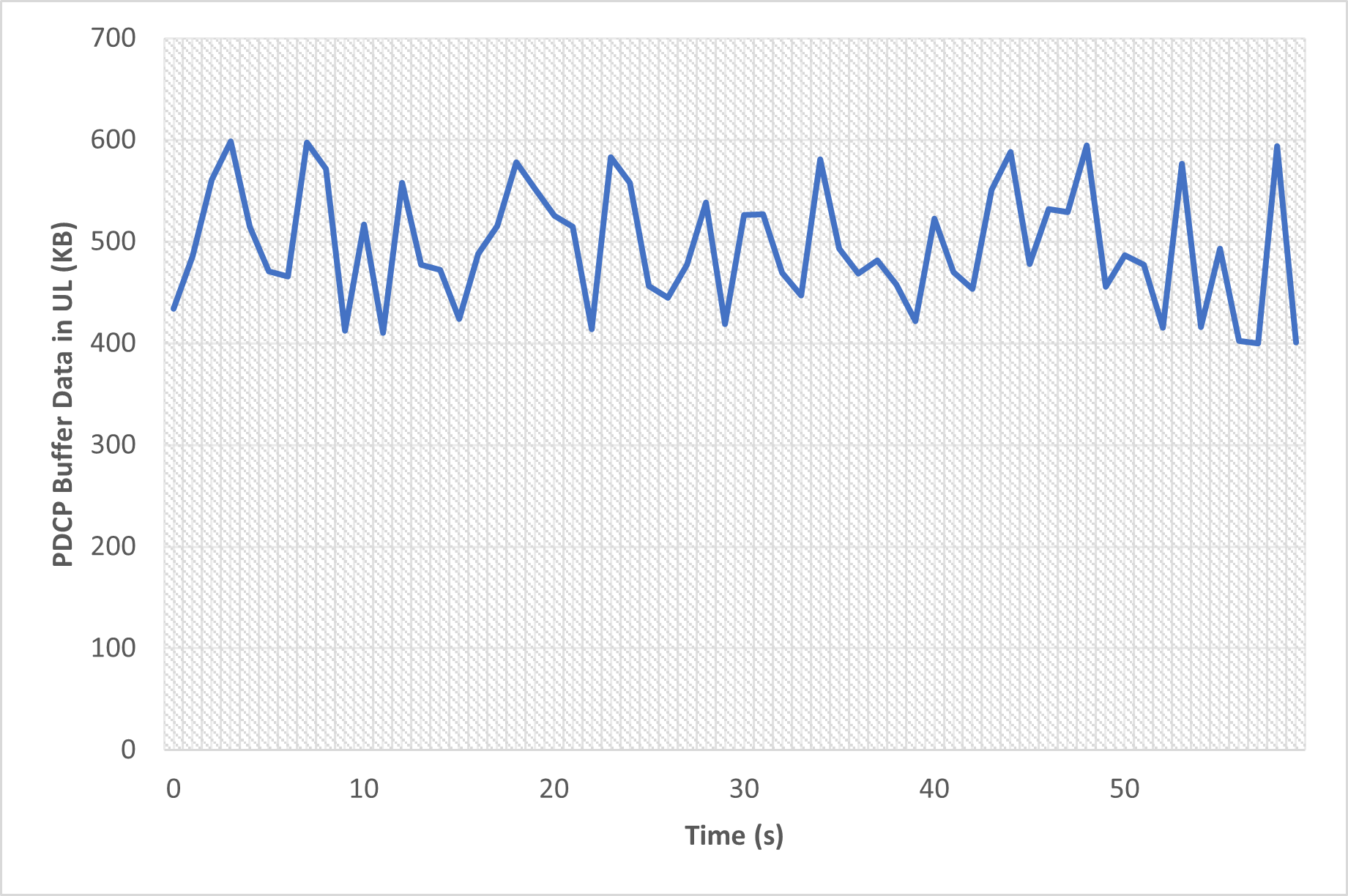}
    \caption{Total PDCP Bytes received by the en-gNB in the uplink from all UEs.}
    \label{fig:KPIMON_plot}
\end{figure}

Fig. \ref{fig:KPIMON_plot} shows the data received in Bytes by the en-gNB from all the UEs connected to it for a particular QCI. In this scenario we have considered two UEs which are transmitting uplink data at a rate of 4 Mbps and 7 Mbps, respectively. Each radio, i.e. the eNB/gNB and each of the two UEs, employ an X310 USRP with its compute node. The KPIMON SM collects the total PDCP Bytes for the entire session while the UE is connected to the en-gNB. The periodic event trigger timer is set to 1 s. 

\subsection{RAN Slicing xApp}
In the previous section we described the KPIMON xApp which made use of the REPORT service. In this section, we demonstrate the working of the RAN Slicing xApp to combine two services to control and change the RAN parameters from the near-RT RIC. The RAN Slicing xApp is extended from POWDER~\cite{Johnson21NexRAN} and integrated into our 5G NSA Testbed. The SM supports the REPORT and CONTROL services. The REPORT service is used to get metrics from the RAN and the CONTROL service is used to vary RAN paramters based on the metrics collected. While the REPORT service is configured the same way as KPIMON xApp with a timer functioning as an event trigger, the CONTROL service acts asynchronously and is initiated by the xApp in the near-RT RIC. 

This xApp configures the en-gNB into different slices and allocates resources to these slices through asynchronous E2 Control Messages. Here a slices is the bandwidth that an operator gets to serve its users, where all operators share a cell. Specifically, we assume three operators, each with one active user. Each UE can thus leverage only the resources available in the network slice allocated to its operator. 

\begin{figure}[t]
    \centering
    \includegraphics[width=0.49\textwidth]{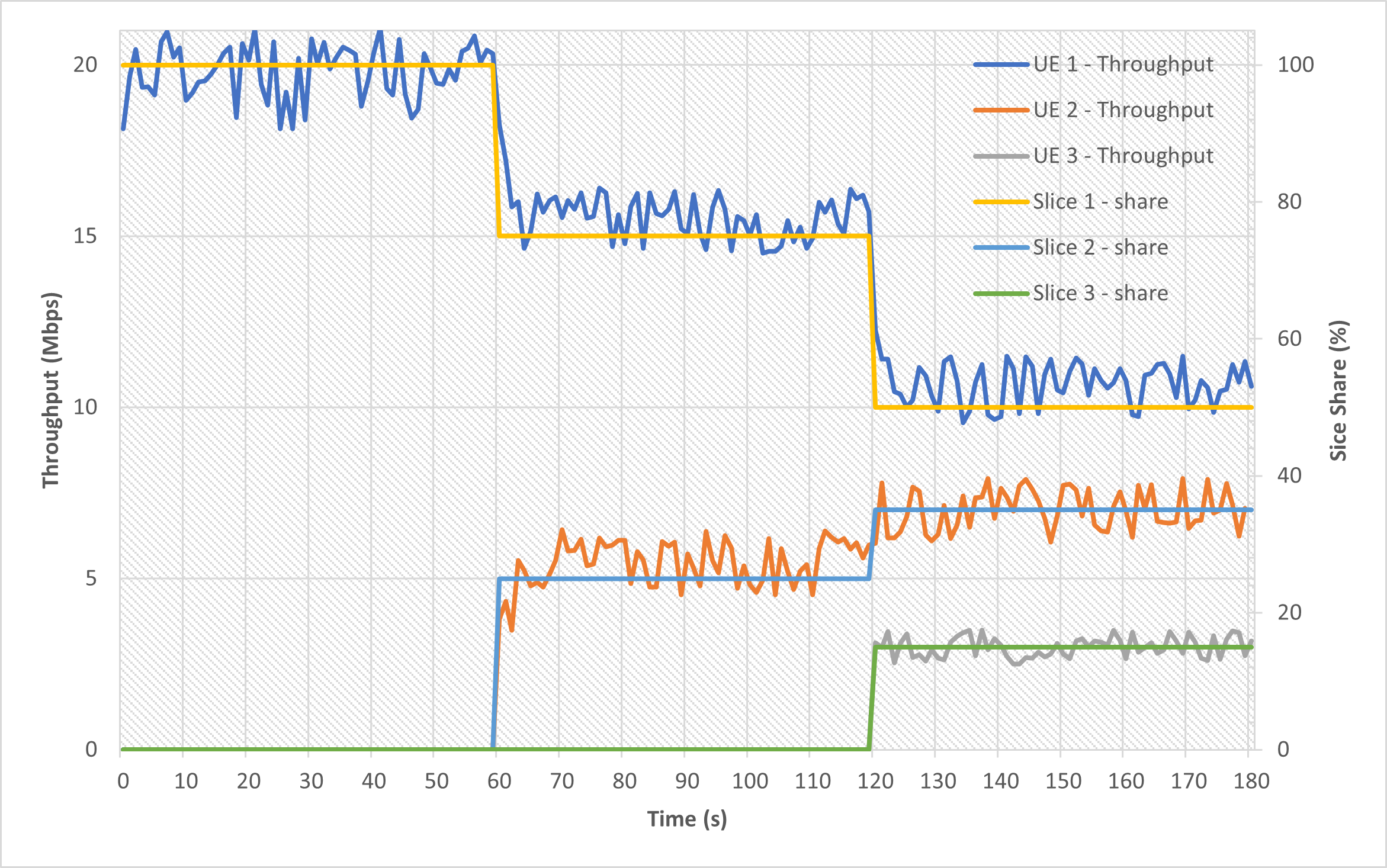}
    \caption{UE bandwidth variation when associated with different slices.}
    \label{fig:RAN_slicing}
\end{figure}

We use one X310 USRP for the base station and one B210 for each of the three UEs, one per operator/slice. Fig. \ref{fig:RAN_slicing} shows the downlink throughput performance of the UEs associated to three different slices with different resource allocations. The shared resource is the 10 MHz LTE cell that is sliced among three operators. A single operator gets the entire system bandwidth for the duration of at least one subframe, after which the cell may be allocated to another operator. This time slicing occurs in a round-robin fashion for the slices that have non-zero resources allocated to them. 
Initially, for the first 60 seconds, slice 1 has access to all the resources, meaning that Slice 1 uses all the subframes for the data transfer between the RAN and the associated UE whereas the other slices have no available subframe resources. After 60 s, Slice 2 is allocated 25\% of the resources and Slice 1 retains the remaining 75\%. As expected, the throughput of the UE drops. After 120 s, the share for Slice 2 increases to 35\% while Slice 1's share is further reduced to 50\%. The remaining resources (15\%) are allocated to Slice 3. The effect of this is visible in the throughput plot. Changing the subframe allocation for each of the slices is also handled through the Control Messages. In parallel, the E2 Indication messages with information on the per slice measurements are reported to the RIC by the E2 Node.      

\section{Conclusion}
\vspace{-1 mm}
\label{sec:conclusion}

The goal of this paper has been to introduce the building blocks of an O-RAN research testbed which uses exclusively COTS hardware and free and open-source software. 
We have therefore evaluated the hardware design options and provided a detailed overview of the 
software suites with particular attention to the near-RT RIC deployment and the E2 interface implementation. The E2 interface is the foundation for the effective inter-working of the various RAN protocol and control processes. 
Our testbed is meant to serve as a blueprint for developing similar laboratory and large-scale O-RAN testbeds to enable 6G research and development in laboratory and production-like setting. To that effect, we have open sourced the software and detailed documentation. 

\section*{Acknowledgement}
This work was supported in part by the National Science Foundation under grant numbers 2120411 and 2120442.
 
\bibliographystyle{IEEEtran}
\bibliography{main.bib}

\section*{Biographies}
\footnotesize

\vspace{0.2cm}
\noindent
\textbf{Pratheek S. Upadhyaya} (pratheek@vt.edu)
is a PhD student in the Department of Electrical and Computer Engineering at Virginia Polytechnic and State University, Blacksburg, VA, USA. His research interests include O-RAN, Scheduling, Network Slicing, AI/ML applications in wireless communications and spectrum sharing.

\vspace{0.2cm}
\noindent
\textbf{Aly Sabri Abdalla} (asa298@msstate.edu)
is a PhD candidate in the Department of Electrical and Computer Engineering at Mississippi State University, Starkville, MS, USA. His research interests are on scheduling, congestion control and wireless security for vehicular ad-hoc and UAV networks.

\vspace{0.2cm}
\noindent
\textbf{Vuk Marojevic} (vuk.marojevic@msstate.edu) is an associate professor in electrical and computer engineering at Mississippi State University, Starkville, MS, USA. His research interests include resource management, vehicle-to-everything communications and wireless security with application to cellular communications, mission-critical networks, and unmanned aircraft systems.

\vspace{0.2cm}
\noindent
\textbf{Jeffrey H. Reed} (reedjh@vt.edu) is currently the Willis G. Worcester Professor with the Bradley Department of Electrical and Computer Engineering, Virginia Tech.

\vspace{0.2cm}
\noindent
\textbf{Vijay K. Shah} (vshah22@gmu.edu) is an assistant professor of cybersecurity engineering at George Mason University, Farifax, VA, USA. His research interests include 5G/next-G communications, AI/ML, wireless security, and wireless testbed development. 


\end{document}